\documentstyle[12pt]{ioplppt}

\eqnobysec

\begin{document}

\jl{1}

\title{Crossover exponent for piecewise directed walk \\adsorption  on Sierpinski fractals}[Crossover exponent for piecewise directed walk ]

\author{S Elezovi\' c--Had\v zi\' c\dag\ftnote{3}{e-mail: suki@rudjer.ff.bg.ac.yu} and N Vasiljevi\' c\ddag\ftnote{4}{e-mail: nale\_vasilj@yahoo.com}}
\address{\dag Faculty of Physics, University of Belgrade, P.O. Box 368, 11001 Belgrade, Serbia} 

\address{\ddag Institute of Biophysics, Faculty of Medicine, University of Belgrade, 
Vi\v segradska 26/2, 11001 Belgrade, Serbia}

\begin{abstract}
We study the problem of critical adsorption of piecewise directed random walks on a boundary of fractal lattices that belong to the Sierpinski gasket family. By applying the exact real space renormalization group method, we calculate the crossover exponent $\phi$, associated with the number of adsorbed steps, for the complete fractal family. We demonstrate that our results are very close to the results obtained for ordinary self--avoiding walk, and discuss the asymptotic behaviour of $\phi$ at the fractal to Euclidean lattice crossover.
\end{abstract}
\maketitle

\section{Introduction}

It is well known that a long flexible linear polymer in a good solvent with an attractive short--range force between the polymer and the container wall undergoes an adsorption transition \cite{DeGennes,Eisenriegler,DeBell}. A good model for this phenomenon is a self-avoiding random walk (SAW) on some lattice with an adsorbing surface (boundary). In this model each monomer (step) in the bulk has a Boltzmann weight $x$, whereas the interaction with the adsorbing wall is taken into account by assigning an energy $\varepsilon_w<0$ to each monomer that is found at the surface. For temperatures $T$ higher than  the critical temperature $T_a$ of the adsorption, the polymer is in a desorbed phase, and for $T<T_a$ it is in an adsorbed phase. At the adsorption transition the number $M$ of adsorbed monomers scales with the total number $N$ of monomers as
$M\sim N^\phi$, where $\phi$ is the crossover exponent.

The above model has been widely studied on various lattices and via a number of techniques (\cite{DeBell} and references therein). For two-dimensional Euclidean lattices
 exact results were found  through conformal field theory \cite{Cardy,Duplantier,Burkhardt}
and much numerical work was completed using Monte Carlo \cite{Meirovitch,Grassberger}, exact enumeration \cite{Barber,Foster,Guttman,Lookman,Zhao}, transfer matrix  \cite{Guim,Veal}, 
series expansion \cite{Domany} and renormalization group \cite{Kremer,Benhamou} techniques. 
Bouchaud and Vannimenus \cite{Bouchaud}  developed a real space renormalization group (RSRG) approach to study the critical adsorption of  SAW on finitely ramified fractal lattices (which may serve as crude models for real amorphous materials). In addition to some other results, 
these authors found the exact values of the  crossover exponent $\phi$ for the case of two-- and three--dimensional Sierpinski gaskets (SG).

The RSRG method \cite{Bouchaud} was  applied on other fractals, in particular on  other members of the two--dimensional SG fractal family \cite{Bubanja}, in order to find out how the crossover exponent changes when properties of the fractal are systematically changed.  Members of this family can be labeled by an integer $b$ ($2\leq b<\infty$), and for large 
values of $b$ the underlying lattice structure becomes more and more similar to a triangular wedge, while its fractal dimension tends to its Euclidean value 2. Due to the extraordinary
complexity of the underlying combinatorial problem, exact crossover exponents $\phi$ were found
\cite{Bubanja} only for $2\leq b\leq 9$. For larger values of $b$ the Monte Carlo renormalization group (MCRG) method  \cite{Zivic} was developed, and the limit $b\to\infty$ was analyzed within
the finite--size scaling hypothesis \cite{Kumar}, with the conclusion that $\phi$ does not tend to its Euclidean value $\phi_E=1/2$ \cite{Burkhardt}.

In this paper we apply the RSRG method to  calculate the crossover exponent $\phi$ for the previously introduced \cite{Capel,PDW2} nontrivial class of SAW, so--called piecewise directed walk (PDW) on the two--dimensional SG family of fractals. For one type
 (PDW1) of this model we were able to find $\phi$ exactly  for the entire SG family,
whereas in the other case (PDW2) we found exact values of $\phi$ for $2\leq b\leq 1000$.  Comparing our results with the  results obtained for the common SAW model on the same fractal family \cite{Bubanja,Zivic} for $2\leq b \leq 100$ we find that for each $b$ the values 
of $\phi$ for both models are very close. This is not surprising since attractive wall at the fractal boundary brings anisotropy in SAW paths which is similar to the effects of the directedness imposed on random walks in PDW models. In other words, it seems plausible to 
accept that disallowed directions does not significantly contribute to the values of $\phi$ 
for the SAW model (especially for large values of $b$), which means that crossover exponent $\phi$ calculated for the PDW model(s) can be considered  as a good approximation for the 
$\phi$ values for SAW on SG fractals. Then, asymptotic analysis of the exact PDW results can 
also provide good insight into the crossover region $b\to\infty$ for the SAW model, which 
might help in resolving the subtle problem of the limiting behaviour of the critical 
properties of statistical systems on fractals when underlying fractal parameters approach 
the corresponding Euclidean values \cite{Bubanja,Zivic,Kumar,Capel,PDW2,PDW2gama,brazil}.

The paper is organized as follows. In section 2 we recall the definition of the PDW model and present the framework of the RSRG calculation of $\phi$ for the case under study.
In sections 3 and 4 we present our results for $\phi$, together with the analysis of the
asymptotic  behaviour in the limit $b\to\infty$, in the case of PDW1 and PDW2, respectively.
Finally, in section 5 we discuss all our findings and related results obtained by other 
authors.

\section{The models and the crossover exponent $\phi$}

In this section we are going to apply the RSRG method to the PDW adsorption problem on the SG family of fractals. We start with recalling the basic properties of  well--known
SG fractals introduced by Given and Mandelbrot \cite{Given}.
Each member of the SG fractal family (labeled by $b$) can be constructed recursively, starting with an equilateral triangle (generator $G_1(b)$) that contains $b^2$ smaller equilateral triangles. The subsequent fractal stages are constructed self--similarly, by replacing each of the $b(b+1)/2$ upward--oriented small triangles of the initial generator by a new generator. To obtain the $r$th--stage fractal lattice $G_r(b)$, which we shall call the generator of
order $r$, this process of construction has to be repeated $r-1$ times, and the complete fractal is obtained in the limit $r\to\infty$. The fractal dimension of the SG fractal with spatial scaling factor $b$ is given by $d_f=\ln[b(b+1)/2]/\ln b$. In the case under study, it is assumed that one side of the fractal is an impenetrable attractive wall.

The first type -- PDW1 -- of the PDW model is defined \cite{Capel} by requiring that within each generator of any order
of a given SG fractal the self--avoiding walker makes random steps in only three directions: the main one (randomly chosen), and the two neighboring (obtained by rotating the main direction by $\pm \pi/3$). In the case of a generator with $r=1$, one step is related to passing through one physically present unit triangle, whereas in the case of a generator with $r>1$ a step should be related to the coarse--grained walk through the underlying fractal structure. The PDW2 type \cite{PDW2} is defined in a similar way: within each generator of any order four directions, determined by the point of entering a generator, are allowed (see figure 1).

To calculate the critical exponent $\phi$ by means of the RSRG approach, it is necessary to consider three restricted partition functions: $B^{(r)}$, $B^{(r)}_1$, and $B^{(r)}_2$, defined as weighted sums over all corresponding never--starting and never--ending PDWs for the given stage $r$ of the iterative construction of the lattice (see figure 2). The recursive nature of the fractal under consideration implies the following general form of the recursion relations:
\begin{eqnarray}
B'&=&\sum_{k}{\cal B}(k)B^{k}\, ,\nonumber \\
B'_1&=&\sum_{k,l,m}{\cal B}_1(k, l, m)B^{k}B_1^{l}
B_2^{m}\, , \nonumber \\
B'_2&=&\sum_{k,l,m}{\cal B}_2(k,l,m)B^{k}B_1^{l}
B_2^{m}\, ,  \label{eq:rgb}
\end{eqnarray}
where we have used the prime symbol as a superscript for the $(r+1)$th-order functions and no indices for the $r$th-order functions. 
The coefficients ${\cal B}(k)$, ${\cal B}_1(k, l, m)$, and ${\cal B}_2(k, l, m)$ are the numbers 
of ways in which the corresponding part of the PDW, within an $(r+1)$th-stage fractal structure, can be comprised  of the PDWs within the fractal structures of the next lower order.
Then, following the  RSRG procedure \cite{Bouchaud}, one should find the symmetric fixed point 
$B=B_1=B_2=B^*$
of the RSRG transformation (\ref{eq:rgb}), which corresponds to the adsorption--desorption transition at temperature $T=T_a$,  linearize
the RSRG transformation  at that point, and find the relevant eigenvalues. Finally, the crossover exponent $\phi$ is determined through the formula
\begin{equation}
{\phi}={\ln {\lambda_{\phi}}\over {\ln  {\lambda_{\nu}}}}\, ,\label{eq:phi}
\end{equation}
where $\lambda_\nu$ is 
\begin{equation}
\lambda_\nu=\left.{{\partial B'}\over{\partial B}}\right|_{B=B^*}\, , \label{eq:lambdanu}
\end{equation}
and $\lambda_\phi$ is the larger eigenvalue of the matrix 
\begin{equation}
{\bf A}=\left.\left(
\matrix{
{{\partial B'_1}\over{\partial B_1}}&{{\partial B'_1}\over{\partial B_2}}\cr
{{\partial B'_2}\over{\partial B_1}}&{{\partial B'_2}\over{\partial B_2}}\cr}
\right)\right|_{B=B_1=B_2=B^*}\, . \label{eq:a}
\end{equation}

\section{The crossover exponent $\phi$ in the case of PDW1} \label{sec:PDW1}

It was found earlier \cite{Capel} that the  first RSRG equation in (\ref{eq:rgb}) in the case 
of PDW1 for any SG fractal has the following explicit form
\begin{equation}
B'=\sum_{k=0}^{b-1}{1\over{k+1}}{b\choose k}{{b-1}\choose k}B^{b+k}\, , \label{eq:rg1pdw1}
\end{equation}
so that $\lambda_\nu$ (\ref{eq:lambdanu}) is equal to
\[
\lambda_\nu=\sum_{k=0}^{b-1}{{b+k}\over{k+1}}{b\choose k}{{b-1}\choose k}(B^*_b)^{b+k-1}\, , 
\label{eq:lnu}
\]
where $B^*_b$ is the unique positive fixed point of the transformation (\ref{eq:rg1pdw1}).
The equation (\ref{eq:rg1pdw1}) was derived using more general partition functions $X_{b,i}(B)$:
\[
X_{b,i}(B)=\sum_{l} {R_{b,i}(l)B^l}\, , \label{eq:helpX}
\]
where $R_{b,i}(l)$  denotes  number of all possible PDW1 paths 
with $l$ steps in the bulk, starting at the origin $(0,0)$ and ending at the point $(b,i)$ in figure 3, 
assuming that the weight of a step in the bulk is $B$. Similarly, the explicit form of the matrix 
{\bf A} (\ref{eq:a}) can be established by introducing general partition functions $Y_{b,i}$ and $Z_{b,i}$:
\begin{eqnarray*}
  Y_{b,i}(B,B_{1},B_{2})&=&
\sum_{l,m,n}{Q_{b,i}(l,m,n)B^lB_{1}^mB_{2}^n} \, , \nonumber\\
   Z_{b,i}(B,B_{1},B_{2})&=&
\sum_{l,m,n}{S_{b,i}(l,m,n)B^lB_{1}^mB_{2}^n}\, , \label{eq:helpYZ}
\end{eqnarray*}
where $Q_{b,i}(l,m,n)$, and $S_{b,i}(l,m,n)$ denote  numbers of all possible PDW1 paths 
with $l$ steps in the bulk, $m$ steps at the wall, and $n$ steps in the layer adjacent to the wall, again starting at the origin $(0,0)$ and ending at the point $(b,i)$ in figure 3, with the additional assumptions that the weight of a step  at the wall is $B_1$, and in the layer adjacent to the wall -- $B_2$. 
One can observe that for $b=1$ and $b=2$ generating functions $Y_{b,i}(B,B_1,B_2)$ are:
\begin{eqnarray}
Y_{1,0}(B,B_1,B_2)=B_2, \quad &&Y_{2,0}(B,B_1,B_2)=B_2B, \nonumber\\
 Y_{1,1}(B,B_1,B_2)=B_1, \quad &&Y_{2,1}(B,B_1,B_2)=B_2(B+B_1), \nonumber\\
&&Y_{2,2}(B,B_1,B_2)=B_1^2+B_2^2B, \label{eq:b12}
\end{eqnarray}
whereas for $b>2$ (see figure 3(a)) the following relations are valid:
\begin{eqnarray}
\fl Y_{b,0}(B,B_1,B_2)=&B_2B^{b-1}, \nonumber \\
\fl Y_{b,1}(B,B_1,B_2)=&B_2B^{b-2}(B_1+(b-1)B),  \nonumber\\
\fl Y_{b+1,i}(B,B_1,B_2)=&\sum_{k=0}^{i-1}B^{i-k}Y_{b,k}(B,B_1,B_2)+BY_{b,i}(B,B_1,B_2), \, {\mathrm{for}} \quad 2\leq i\leq b-1, \nonumber \\
\fl Y_{b+1,b}(B,B_1,B_2)=&\sum_{k=0}^{b-1}B^{b-k}Y_{b,k}(B,B_1,B_2)+B_2Y_{b,b}(B,B_1,B_2), \nonumber
\\
\fl Y_{b+1,b+1}(B,B_1,B_2)=&\sum_{k=0}^{b-2}B^{b-k}B_2Y_{b,k}(B,B_1,B_2)
+B_1Y_{b,b}(B,B_1,B_2)\, . \label{eq:rec1}
\end{eqnarray}
Starting with the above equations, one can show (see \ref{app:prvi}) that elements $A_{11}$ and 
$A_{12}$ of the matrix {\bf A} (\ref{eq:a}) are equal to:
\begin{equation}
\fl A_{11}(b)=\left.{{\partial B_1'}\over{\partial B_1}}\right|_{*}=
\left.{{\partial Y_{b,b}}\over{\partial B_1}}\right|_{*}=
2\sum_{k=0}^{b-2}{1\over{k+1}}{b\choose k}  {{b-2}\choose k}(B^*_b)^{2b-k-3}\, , \label{eq:a11}
\end{equation}
\begin{eqnarray}
\fl A_{12}(b)=\left.{{\partial B_1'}\over{\partial B_2}}\right|_{*}=
\left.{{\partial Y_{b,b}}\over{\partial B_2}}\right|_{*}=2(B^*_b)^b\left(
\sum_{k=0}^{b-2}{1\over{k+2}}{b\choose {k+1}}{{b-1}\choose {k+1}}(B^*_b)^k \right.\nonumber\\
\lo+\left.\sum_{k=0}^{b-3}{{2k}\over{(k+2)(k+3)}}{b\choose{k+1}}{{b-2}\choose{k+1}}
(B^*_b)^k\right)\, , \label{eq:a12}
\end{eqnarray}
where $*$ denotes that partial derivatives are calculated at the symmetric fixed point 
$B=B_1=B_2=B_b^*$. As one can see on figure 3(b), the restricted partition
function $B_2'$ for the SG fractal labeled by $b$ is
\begin{equation}
B_2'=Z_{b,b}(B,B_1,B_2),
\end{equation}
and
\begin{equation}
Z_{b+1,b+1}(B,B_1,B_2)=\sum_{i=0}^bB_2B_1^iX_{b,b-i}(B)\, ,
\end{equation}
where
\begin{eqnarray}
X_{b,0}(B)&=B^b, \quad {\rm and} \nonumber \\
X_{b,b-i}(B)&=\sum_{k=i}^{b-1}{{i+1}\over{k+1}}{b\choose k}{{b-i-1}\choose{k-i}} B^{b-i+k}, \qquad {\rm for} \quad i<b,
 \label{eq:Z2}
\end{eqnarray}
as was shown in \cite{Capel}. Then,
\begin{equation}
A_{21}(b)=
\left.{{\partial Z_{b,b}}\over{\partial B_1}}\right|_{*}=
\sum_{i=1}^{b-1}iB^iX_{b-1,b-1-i}(B), \label{eq:Z1}
\end{equation}
which can be put (see \ref{app:drugi}) in the following form
\begin{equation}
A_{21}(b)=\sum_{k=1}^{b-1}{2\over{k+1}}{{b-1}\choose k}{b\choose{k-1}}(B^*_b)^{b-1+k}\, .  \label{eq:a21}
\end{equation}
For the last element $A_{22}$ of the matrix {\bf A} (\ref{eq:a}) we have
\[
A_{22}(b)=
\left.{{\partial Z_{b,b}}\over{\partial B_2}}\right|_{*}=\left.
\sum_{i=0}^{b-1} B^iX_{b-1,b-1-i}(B)\right|_{*}\, . 
\]
From (\ref{eq:rec1}), with $B=B_1=B_2$ and $Y_{b,i}(B,B,B)=X_{b,i}(B)$, we get
\begin{equation}
X_{b,b}(B)=\sum_{i=0}^{b-1} B^{i+1}X_{b-1,b-1-i}(B)\, ,
\end{equation}
so that $A_{22}(b)$ is equal to:
\begin{equation}
A_{22}(b)=\left.{1\over B}X_{b,b}(B)\right|_{*}.  \label{eq:ZZZ2}
\end{equation}
Since $B^*_b$ is the unique positive fixed point of the RSRG equation $B'=B$, where $B'=X_{b,b}(B)$, from (\ref{eq:ZZZ2})
directly follows that
\begin{equation}
A_{22}(b)=1.
\end{equation}

The explicit values of fixed points $B^*_b$ and eigenvalues $\lambda_\nu$ were found in \cite{Capel} for $2\leq b\leq 1000$ and for several larger values of $b$. Knowing $B^*_b$, it is a matter of straightforward calculation to find elements $A_{11}$, 
$A_{12}$, and $A_{21}$(equations (\ref{eq:a11}),(\ref{eq:a12}), and (\ref{eq:a21})), then the larger eigenvalue $\lambda_\phi$ of the corresponding matrix ${\bf A}$, and finally the crossover exponent $\phi$ (\ref{eq:phi}). 
In figure 4 we present the complete sequence of  $\phi$--values up to $b=1000$, and also for $b=5000$ and $b=10 000$.
These values are compared with the theoretical values obtained via the asymptotic  analysis for large $b$, which we are going to present now.

For large values of $b$, introducing new variable $k=bx$ the summation over $k$ in equation (\ref{eq:a11}) can be replaced by integration over $x$.
Then, by the use of the Stirling formula for large $n$, that is
\[
\ln n!=\left(n+{1\over 2}\right)\ln n-n+{1\over 2}\ln 2\pi \, , 
\]
in binomial coefficients, element $A_{11}(b)$ can be written in the form
\begin{equation}
A_{11}(b)={2\over{(B^*_b)^3}}{1\over{2\pi b}}\int_0^1 {{1-x}\over{x^2}}{e}^{-2bh(x)}\, \d x \quad , \label{eq:inta11}
\end{equation}
where $h(x)$ is
\[
h(x)=x\ln x+(1-x)\ln(1-x)-{1\over 2}(2-x)\ln B^*_b\, .
\]
Since $b$ is considered to be very large the integral in equation (\ref{eq:inta11}) is treated here by the use of the steepest-descent method expanding the integrand around the minimum of $h(x)$, which occurs at $x_0=1/(1+\sqrt{B^*_b})$, so that $A_{11}(b)$ can be expressed as
\begin{equation}
A_{11}(b)={2\over{(B^*_b)^3}}{1\over{2\pi b}}{{1-x_0}\over{x_0^2}}\sqrt{
{\pi\over{bh''(x_0)}}} \e^{-2bh(x_0)}\, . \label{eq:enda11}
\end{equation}
By similar arguments, in ref. \cite{Capel} was shown that the first equation in RSRG transformation (\ref{eq:rgb}) can be put in the form
\begin{equation}
B'={1\over{2\pi b}}{1\over{x_1^2}}\e^{-2bg(x_1)}\sqrt{
{\pi\over{bg''(x_1)}}} \, , \label{eq:capelrg}
\end{equation}
for large $b$, 
where
\[
g(x)=x\ln x+(1-x)\ln (1-x)-{1\over 2}(1+x)\ln B 
\]
has its minimum at $x_1=\sqrt{B}/(1+\sqrt{B})$. It is trivial to show that $h(x_0)=g(x_1)$ and 
$h''(x_0)=g''(x_1)$, for $B=B^*_b$. Then, by direct comparison of equations (\ref{eq:enda11}) and (\ref{eq:capelrg}), one can see that  
\begin{equation}
A_{11}(b)={2\over{\sqrt{B^*_b}(1+\sqrt{B^*_b})}}= 2-{3\over 4}(3+\sqrt 5){{\ln b}\over b} +
O\left({1\over b}\right)\, ,
\end{equation}
since $B'=B^*_b$ if $B=B^*_b$, and
\begin{equation}
B^*_b= {{3-\sqrt 5}\over 2}+{3\over 4}{{5-\sqrt 5}\over 5}{{\ln b}\over b}+O\left({1\over b}\right) 
\end{equation}
(which was also shown in ref. \cite{Capel}). 

In a quite similar way from equation (\ref{eq:a12})  follows
\begin{eqnarray}
\fl A_{12}(b)=2\left(1-(B^*_b)^{b-1}+{1\over{\pi bB^*_b}}\int_0^1{{1-x}\over{x^2}}\e^{-2bg(x)}\, \d x\right)\approx{4\over{1+\sqrt{B^*_b}}} \nonumber\\
\lo= 2\sqrt 5-{3\over{10}}(3\sqrt 5-5){{\ln b}\over b}+O\left({1\over b}\right)\, , 
\end{eqnarray}
 whereas from equation (\ref{eq:a21}) one gets
\begin{equation}
\fl A_{21}(b)={1\over{\pi b B^*_b}}\int_0^1 {{\d x}\over{x(1-x)}}
\e^{-2bg(x)}=2\sqrt{B^*_b}
=\sqrt 5-1+{{\sqrt 5}\over{10}} {{\ln b}\over b}+O\left({1\over b}\right)\, .
\end{equation}
Then, the largest eigenvalue $\lambda_\phi$ of the matrix $\bf A$ (\ref{eq:a})
 is  approximately equal to
\begin{equation}
\lambda_\phi(b)=C_1-C_2 {{\ln b}\over b}\, , \label{eq:lfi}
\end{equation}
 for $b\gg 1$, where
\[
C_1=\lim_{b\to\infty}\lambda_\phi(b)={1\over 2}\left(
3+(41-8\sqrt 5)^{1/2}\right)=3.9037... \, ,
\]
\[
C_2={1\over{40}}
\left(
(41 - 8\,\sqrt{5})^{1/2}
( 5 - \sqrt{5})  +15
( 3 + \sqrt{5} ) \right)=2.2957...\, . 
\]

Using asymptotic behaviour of $\lambda_\nu$, which has also been treated in \cite{Capel}, i.e.
\begin{equation}
\lambda_\nu(b)={{5-\sqrt 5}\over 2}b+{3\over{40}}(3\sqrt 5-5)\ln b+\cdots\, ,
\end{equation}
together with (\ref{eq:lfi}) and (\ref{eq:phi}), we finally get the following expression for the crossover exponent
\begin{equation}
\phi(b)={{\ln C_1}\over{\ln b}}\left(1-{
{\ln{{5-\sqrt 5}\over 2}}
\over
{\ln b}
}
+\cdots\right)\, . \label{eq:asymphi1}
\end{equation}
The exponent $\phi$ given by (\ref{eq:asymphi1}) is represented as the dashed curve in 
figure 4, and one can see that agreement with the exact results is quite satisfactory.
In the same figure, we have presented the lower $\phi_l$ and the upper $\phi_u$ bounds
\begin{equation}
\phi_l=1-(d_f-d_s)\nu\quad , \qquad \phi_u={{d_s}\over{d_f}} \label{eq:bounds}
\end{equation}
for the crossover exponent $\phi$, established in a heuristic way by Bouchaud and Vannimenus in
\cite{Bouchaud}. In the above bounds $d_s$ is the fractal dimension of the adsorbing surface,
i.e. $d_s=1$ for our model, whereas $\nu$, the critical exponent of the end-to-end distance of 
walk in the bulk, is equal to $\ln b/\ln\lambda_\nu$. One can observe that our results satisfy
the condition $\phi_l\leq\phi\leq\phi_u$ for any $b$, which is not the case for the SAW model
with no constraints (see figure 3 in \cite{Zivic}).

\section{The crossover exponent $\phi$ in the case of PDW2 \label{sec:pdw2}}

For the PDW2 model in the bulk it is enough to say that within each generator of any order of 
a given fractal the walker makes random steps represented only by the set of four  vectors which corresponds to the place of his entering the generator (see figure 1(b)), and it is  not relevant
which end of the walk is starting point. This is not the case in the vicinity of the adsorbing wall. See, for instance, two configurations in figure 5, which are symmetric under change of orientation of the walk. If the wall is not present both of them represent PDW2 walks, and contribute the RG equation for the $B^{(r+1)}$ parameter with the same term 
$(B^{(r)})^6$. When the wall is present these two configurations contribute the $B_2^{(r+1)}$ parameter, first with the term $(B^{(r)})^3(B_2^{(r)})^3$, and second with $(B^{(r)})^3(B_1^{(r)})^2B_2^{(r)}$, so it is obvious that the PDW2 definition in the case of adsorption should be completed to avoid such ambiguities. In that sense, two types of PDW2 
walks are possible. We shall denote them by abbreviations PDW2a and PDW2b, and describe in the following way:
\begin{itemize}  
\item PDW2a model -- parts of the PDW2a walks that traverse generators lying on the adsorbing wall are oriented in such a way that their entrance point is always on the wall (first example in figure 5);
\item PDW2b model -- entrance point of the PDW2b walks that traverse generators lying on the adsorbing wall is always in the bulk (second example in figure 5).
\end{itemize}

As in the case of the PDW1 model, it is useful to introduce analogous partition functions $X_{b,i}(B)$, $Y_{b,i}(B,B_1,B_2)$, and $Z_{b,i}(B,B_1,B_2)$ for both PDW2 models.
Functions $X_{b,i}(B)$ were already introduced in reference \cite{PDW2}, but no closed form
expression for them were found. Nevertheless, these functions, as well as their derivatives, 
obey certain recursive relations, by the use of which one can numerically find fixed points
$B^*_b$ of the RSRG transformation $B'=B$ (\ref{eq:rgb}) and the eigenvalues $\lambda_\nu(b)$
(\ref{eq:lambdanu}), for very large values of the SG fractal parameter $b$ \cite{PDW2,PDW2gama}.
Here we shall describe the algorithm for numerical determining the elements of the matrix {\bf A}  (\ref{eq:a}). 

For both versions of the PDW2 model the RSRG transformation of the function $B_1'$, for any $b$, can be expressed as $B_1'=Y_{b,b}(B,B_1,B_2)$, so that elements $A_{11}(b)$ and $A_{12}(b)$ of the matrix {\bf A}  (\ref{eq:a}) are equal to
\begin{equation}
A_{11}(b)=Y_{b,b}^1(B_b^*), \, A_{12}(b)=Y_{b,b}^2(B_b^*), \label{eq:vrsta1}
\end{equation}
where we have used short labeling
\begin{equation}
Y_{b,i}^1(B)\mathop{=}^{\rm def}\left.{{\partial Y_{b,i}}\over{\partial B_1}}\right|_{B=B_1=B_2},\,
Y_{b,i}^2(B)\mathop{=}^{\rm def}\left.{{\partial Y_{b,i}}\over{\partial B_2}}\right|_{B=B_1=B_2}
\end{equation}
It is not difficult to check that for $b=1$ and $b=2$ generating functions $Y_{b,i}(B,B_1,B_2)$ are:
\begin{eqnarray*}
Y_{1,0}(B,B_1,B_2)=B_2, \quad &&Y_{2,0}(B,B_1,B_2)=B_2B(1+B) , \nonumber\\
 Y_{1,1}(B,B_1,B_2)=B_1, \quad &&Y_{2,1}(B,B_1,B_2)=B_2(B+B_1), \nonumber\\
&&Y_{2,2}(B,B_1,B_2)=B_1^2+B_2^2B, 
\end{eqnarray*}
whereas for $b>2$ the following relations are valid:
\begin{eqnarray}
\fl Y_{b+1,0}(B,B_1,B_2)=&\sum_{k=0}^{b-1}B^{k+1}Y_{b,k}+B^bB_2Y_{b,b}\, , \nonumber \\
\fl Y_{b+1,1}(B,B_1,B_2)=&BY_{b,0}+\sum_{k=1}^{b-1}B^k Y_{b,k}+
B^{b-1}B_2Y_{b,b}\, ,  \nonumber\\
\fl Y_{b+1,i}(B,B_1,B_2)=&\sum_{k=0}^{i-1}B^{i-k}Y_{b,k}+\sum_{k=i}^{b-1}B^{k-i+1}Y_{b,i}
+B^{b-i}B_2Y_{b,b}\, , \quad {\mathrm{for}} \quad 2\leq i\leq b-1, \nonumber \\
\fl Y_{b+1,b}(B,B_1,B_2)=&\sum_{k=0}^{b-1}B^{b-k}Y_{b,k}+B_2Y_{b,b}\, , \nonumber
\\
\fl Y_{b+1,b+1}(B,B_1,B_2)=&\sum_{k=0}^{b-1}B^{b-k}B_2Y_{b,k}+B_1Y_{b,b}\, , \label{eq:pdw2rec1}
\end{eqnarray}
as one can see on figure 6(a). From these equations one obtains recursive relations for the derivatives $Y_{b,i}^1(B)$ and  $Y_{b,i}^2(B)$:
\begin{eqnarray}
\fl Y_{b+1,0}^1(B)&=\sum_{k=0}^{b}B^{k+1}Y_{b,k}^1(B)\, , \nonumber \\
\fl Y_{b+1,1}^1(B)&=BY_{b,0}^1(B)+\sum_{k=1}^{b-1}B^k Y_{b,k}^1(B)\, ,  \nonumber\\
\fl Y_{b+1,i}^1(B)&=\sum_{k=0}^{i-1}B^{i-k}Y_{b,k}^1(B)+\sum_{k=i}^{b}B^{k-i+1}Y_{b,i}^1(B)\, , \quad {\mathrm{for}} \quad 2\leq i\leq b-1, \nonumber \\
\fl Y_{b+1,b}^1(B)&=\sum_{k=0}^{b-1}B^{b-k}Y_{b,k}^1(B)+BY_{b,b}^1(B)\, , \nonumber
\\
\fl Y_{b+1,b+1}^1(B)&=\sum_{k=0}^{b}B^{b-k+1}Y_{b,k}^1(B)+X_{b,b}(B)\, , \label{eq:pdw2rec2}
\end{eqnarray}
and
\begin{eqnarray}
\fl Y_{b+1,0}^2(B)&=\sum_{k=0}^{b}B^{k+1}Y_{b,k}^2(B)+B^bX_{b,b}(B)\, , \nonumber \\
\fl Y_{b+1,1}^2(B)&=BY_{b,0}^2(B)+\sum_{k=1}^{b-1}B^k Y_{b,k}^2(B)+
B^{b-1}X_{b,b}(B)\, ,  \nonumber\\
\fl Y_{b+1,i}^2(B)&=\sum_{k=0}^{i-1}B^{i-k}Y_{b,k}^2(B)+\sum_{k=i}^{b}B^{k-i+1}Y_{b,i}^2(B)
+B^{b-i}X_{b,b}(B)\, , \quad {\mathrm{for}} \quad 2\leq i\leq b-1, \nonumber \\
\fl Y_{b+1,b}^2(B)&=\sum_{k=0}^{b-1}B^{b-k}Y_{b,k}^2(B)+BY_{b,b}^2(B)+X_{b,b}(B)\, , \nonumber
\\
\fl Y_{b+1,b+1}^2(B)&=\sum_{k=0}^{b}B^{b-k+1}Y_{b,k}^2(B)+\sum_{k=0}^{b-1}B^{b-k}X_{b,k}(B)\, . \label{eq:pdw2rec3}
\end{eqnarray}
Then, starting from the initial values 
\begin{eqnarray*}
X_{1,0}(B)&=Y_{1,0}(B,B,B)=B\, , \quad X_{1,1}(B)=Y_{1,1}(B,B,B)=B\, , \nonumber\\
Y_{1,0}^1(B)&=0\, , \quad Y_{1,1}^1(B)=1\, , \quad
Y_{1,0}^2(B)=1\, , \quad Y_{1,1}^2(B)=0\, , 
\end{eqnarray*}
with $B=B^*_b$, for any $b$ one can iterate relations (\ref{eq:pdw2rec1}), (\ref{eq:pdw2rec2})
and (\ref{eq:pdw2rec3}), and obtain values of the elements $A_{11}(b)$ and $A_{12}(b)$ 
(\ref{eq:vrsta1}).

One can observe that for the PDW2a model
\begin{equation}
B_2'=Y_{b,0}(B,B_1,B_2)\, ,
\end{equation}
so that
 elements $A_{21}(b)$ and $A_{22}(b)$ are equal to
\begin{equation}
A_{21}(b)=Y_{b,0}^1(B_b^*)\, , \quad A_{22}(b)=Y_{b,0}^2(B_b^*)\, .
\end{equation}
which can also be obtained by the use of recursion relations (\ref{eq:pdw2rec1}), (\ref{eq:pdw2rec2}) and (\ref{eq:pdw2rec3}). 

In the case of the PDW2b model
\begin{equation}
B_2'=Z_{b,0}(B,B_1,B_2)=Z_{b,b}(B,B_1,B_2)
\end{equation}
and
\begin{equation}
A_{21}(b)=\left.{{\partial Z_{b,0}}\over{\partial B_1}}\right|_*\, , \quad
A_{22}(b)=\left.{{\partial Z_{b,0}}\over{\partial B_2}}\right|_*\, .
\label{eq:vrsta2}
\end{equation}
Function $Z_{b,0}(B,B_1,B_2)$ satisfies the relation
\begin{equation}
Z_{b,0}(B,B_1,B_2)=B_2\sum_{i=0}^{b-1}B_1^{i}X_{b-1,i}(B)\, ,
\end{equation}
as can bee seen from figure 6(b). Then, from (\ref{eq:vrsta2}) directly follows
\begin{equation}
A_{21}(b)=\sum_{i=1}^bi(B^*_b)^iX_{b-1,i}(B_b^*)\, ,
\end{equation}
together with
\begin{equation}
A_{22}(b)=\sum_{i=0}^{b-1}B^iX_{b-1,i}(B_b^*)   \, .
\end{equation}
The form of the $A_{22}(b)$ can be further simplified, since the sum on the right side of the
last equation is equal to $X_{b,0}(B_b^*)/B_b^*=X_{b,b}(B_b^*)/B_b^*$, which is direct consequence 
of relations (\ref{eq:pdw2rec1}) with $B=B_1=B_2$. Finally, we obtain
\begin{equation}
A_{22}(b)=1\, ,
\end{equation}
since 
$X_{b,b}(B_b^*)=B^*_b$.

In figure 7(a) we present larger eigenvalue $\lambda_\phi$ of the matrix {\bf A} (4) for SG fractals with $b\leq 1000$ for both PDW2 models (elements of the matrix {\bf A} were
calculated by the use of the previously found \cite{PDW2,PDW2gama} fixed points $B^*_b$). It seems
that $\lambda_\phi$  tends to some finite constant when $b\to\infty$, for both cases, which was also supported by numerical analysis of the obtained data. Furthermore, previous numerical analysis of PDW2 $\lambda_\nu$ eigenvalues \cite{PDW2,PDW2gama} revealed that $\lambda_\nu\sim b$ for large $b$, which means that crossover exponent $\phi$ (\ref{eq:phi}) tends to 0 as const$/\ln b$.  That conclusion is in accord with figure 7(b), where calculated values of $\phi$  are depicted as functions of $1/\ln b$. In the same figure one can also see that, as in the PDW1 case, Bouchaud--Vannimenus boundaries (\ref{eq:bounds}) are obeyed.

\section{Discussion}

For the sake of a better review of global behaviour of  $\phi$ as function of the fractal parameter $b$, when number of allowed directions grows from three (PDW1 model) to six (SAW model), in figure 8 we present our exact results for all PDW models, together with the exact ($2\leq b\leq 9$, \cite{Bubanja}) and available Monte Carlo ($9< b\leq 100$, \cite{Zivic}) SAW results. For all considered $b$, crossover exponents $\phi$ are very close
for all these models, being monotonically decreasing functions of $b$. More precisely, for $12<b<15$ values of $\phi$ for PDW1, PDW2b and SAW are almost the same (see the inset in figure 8), whereas PDW2a values slightly departure from them. For smaller values of $b$ ($b<10$) inequality $\phi_{\mathrm PDW2a}\leq \phi_{\mathrm PDW1}\leq \phi_{\mathrm SAW}\leq \phi_{\mathrm PDW2b}$ holds, whereas for larger $b$ ($b>15$), one can observe  $\phi_{\mathrm PDW2a}<\phi_{\mathrm PDW2b}<\phi_{\mathrm PDW1}$. As $b$ becomes larger $\phi$ values for all three PDW models become closer, indicating that main contribution to the value of $\phi$
comes from walks with steps in three directions allowed in the PDW1 model. The fact that 
PDW2a values of $\phi$ are always smaller than PDW2b values means that the fourth direction
in the PDW2a model (which is forbidden to the PDW1 walker) moves the walker more
efficiently from the adsorbing wall, than the corresponding direction in the PDW2b model (see figure 5). Good illustration for the interplay between the effects of increasing $b$ 
and changing the number of step directions is the relationship between PDW1 and PDW2b data:
for smaller $b$ values, i.e. for smaller fractal generators, there is not enough space for the
fourth direction to move the walker away from the adsorbing wall (so 
$\phi_{\mathrm PDW2b}>\phi_{\mathrm PDW1}$), which is not the case in larger generators ($\phi_{\mathrm PDW2b}<\phi_{\mathrm PDW1}$). 

Exponent $\phi$ for SAW model decreases with $b$ more rapidly than for any other model considered here. For smaller $b$ exponent $\phi_{\mathrm SAW}$ is close to the corresponding PDW1 and PDW2b values, then, as $b$ grows, $\phi_{\mathrm SAW}$ approaches PDW2a values, and starting from $b=80$  drops even bellow them. One could expect that this trend will continue for $b>100$, since increasing of $b$ makes relatively more steps available in the bulk for the walk to spread, than on the surface. However, according to finite-size scaling prediction \cite{Kumar}, $\phi$ should tend to $1/4$ when $b\to\infty$, which means that $\phi_{\mathrm SAW}$ should have minimum (at least one) for some $b>100$, and the same conclusion holds if one accepts the hypothesis that $\phi_{\mathrm SAW}$ tends to its Euclidean value $\phi_E=1/2$. There is still no firm evidence that either of these assumptions is correct. On the other hand, in this paper we have analytically proven that $\phi_{\mathrm PDW1}\to 0$, and numerically shown that $\phi_{\mathrm PDW2}\to 0$, whereas the value of $\phi$ for directed walk model on Euclidean lattices \cite{Privman} is equal to 1/2 (see Appendix C), as for SAW model. Since PDW walks become directed  in the limit $b\to\infty$, these results definitely mean that crossover exponent $\phi$ on fractals does not necessarily approach its Euclidean value and, also,  give support to our expectation that $\phi_{\mathrm SAW}\to 0$. Whether this expectation is correct or not, we believe that PDW models, being tractable for very large values of $b$,  may serve as excellent toy models for checking the validity of the finite-size scaling hypothesis and the RGMC results for random walk models on fractals.

At the end one can pose the question about the dependence of the crossover exponent $\phi$ on 
the fractal properties. There is no simple function of a limited set of the fractal characteristics with which $\phi$ can be described, however, on the basis of equation (\ref{eq:asymphi1}) and numerical analysis of the PDW2 data, we can say that behaviour of
$\phi$ at the fractal to Euclidean lattice crossover is governed by the factor $1/\ln b\sim 2-d_f$, i.e. with the fractal dimension. Similar behaviour was found   for the PDW critical exponent $\nu$ \cite{Capel}, which approaches its Euclidean value $\nu=1$ with the correction term proportional to $1/\ln b$. It is interesting to note here that direct consequence of these two findings -- relation $\phi\sim 1-\nu$, was also predicted by the finite-size scaling method for the SAW model \cite{Kumar}. At the same time, the lower Bouchaud--Vannimenus boundary $\phi_l$
(\ref{eq:bounds}) behaves as $1-\nu$ for large $b$, and for all PDW models considered here exponent $\phi$ approaches $\phi_l$ when $b\to\infty$. The same holds for SAW model on $n$--simplex fractals \cite{simplex}, i.e. $\phi \approx \phi_l$ when $n\to\infty$, and in the
case of simple random walks on $d$--dimensional Sierpinski gaskets $\phi$ is exactly equal to
the lower boundary, for any $d$ \cite{Borjan}. In contrast to these cases,  Bouchaud--Vannimenus boundaries are violated for the SAW on Sierpinski fractals \cite{Zivic}, as well as in some other models \cite{Foster,Vlada,Diehl,Vanderzande}. Further comparative analysis of all these cases should be done in order to make the only known limits for the crossover exponent $\phi$
more precise, which will be a matter for future study.

\ack

This work has been partially supported by the Serbian Science Foundation under Project "Condensed matter physics and new materials".

\Figures

\begin{figure}
\caption{The fractal structure $G_2(3)$ of the $b=3$ SG fractal at the second stage of construction, with examples of (a) PDW1 and (b) PDW2 paths. Arrows represent the allowed
step directions of the corresponding  coarse-grained PDW paths. The gray area at the
basis of $G_2(3)$  represents the adsorption wall. }
\end{figure}
\begin{figure}
\caption{ Schematic representation of the three restricted partition functions, for an 
$r$th stage fractal structure $G_r(b)$ used in the calculation of the PDW crossover 
exponent $\phi$. For example, $B_2^{(r)}$ represents the PDW path that starts at the $G_r(b)$ left vertex that lies on the adsorption wall, and exits the $G_r(b)$ at the upper vertex that lies in the bulk. The interior details of the $G_r(b)$ are not shown (they are manifested by the wiggles of the PDW paths). }
\end{figure}
\begin{figure}
\caption{Illustration of the PDW1 recursion relations (3.3) and (3.7) for the $b+1=6$ SG fractal. 
The solid straight lines  represent the contribution to $Y_{6,3}$ (a) and $Z_{6,6}$ (b) 
from the PDW1 paths via the legitimate paths represented by the solid curves. Arrows represent the possible directions of steps. }
\end{figure}
\begin{figure}
\caption{Crossover exponent $\phi$ for SG fractals with $2\leq b\leq 1000$, $b=5000$, and
$b=10 000$ in the case of PDW1 model ($\triangle$). The dashed line is the theoretical estimate (3.23) of $\phi$  obtained by the asymptotic analysis. The two solid curves represent the bounds
(3.24) for $\phi$.  }
\end{figure}
\begin{figure}
\caption{Examples for PDW2a and PDW2b configurations on the generator of the b=3 SG fractal. In the PDW2a case starting point {\bf A} of the walk is on the adsorbing wall, whereas the PDW2b starting point {\bf B} is in the bulk. }
\end{figure}
\begin{figure}
\caption{Illustration of the PDW2 recursion relations (4.3) and (4.10). The solid straight lines represent the contribution to $Y_{6,3}$ (a) and $Z_{6,0}$ (b) from the PDW2 paths via the legitimate paths represented by the solid curves. Arrows represent the possible directions of steps.}
\end{figure}
\begin{figure}
\caption{(a) The larger eigenvalue $\lambda_\phi$ of the matrix {\bf A} (2.4) for  PDW2a ($\Diamond$) and PDW2b (\opensqr) model for SG fractals with $5\leq b\leq 1000$. 
(b) Crossover exponent $\phi$ for SG fractals with $2\leq b\leq 1000$ in the case of 
PDW2a ($\Diamond$) and PDW2b (\opensqr) model. The two solid lines represent the bounds
(3.24) for $\phi$ for both PDW2 models.} 
\end{figure}
\begin{figure}
\caption{Crossover exponent $\phi$ as function of $1/\ln b$ for the PDW1 ($\triangle$), 
PDW2a ($\Diamond$), PDW2b  (\opensqr) and SAW ($\circ$, data taken from references [20, 21]) models on SG fractals. The full line is the theoretical estimate (3.23) obtained by the asymptotic analysis of the PDW1 data in the limit $b\to\infty$. The inset represents magnified  region $9<b<92$.}
\end{figure}

\appendix

\section{\label{app:prvi}}
In this appendix we present the derivation of equation (\ref{eq:a11}).
We want to prove that 
\begin{eqnarray}
Y_{b,0}^1(B)&=&0\, , \quad Y_{b,1}^1(B)=B^{b-1}\, ,\nonumber \\
Y_{b,i}^1(B)&=&\sum_{k=0}^{i-2}{{b-i+2}\over{k+1}}{b\choose k}{{i-2}\choose k}B^{b+i-k-3}, \quad {\mathrm for}\quad i\geq 2 \, , \label{eq:formula1}
\end{eqnarray}
where
\begin{equation}
Y_{b,i}^1(B)\mathop{=}^{\mathrm def}\left.{{\partial Y_{b,i}}\over{\partial B_1}}\right|_{B=B_1=B_2}
\end{equation}
The proof of equations (\ref{eq:formula1}) proceeds in four steps (i)--(iv):

\noindent
(i) Taking into account equations (\ref{eq:b12}) one can easily verify the relations for $b=1$ and $b=2$ 
\begin{eqnarray}
Y_{1,0}^1(B)&=0\, , \quad Y_{1,1}^1(B)=1\nonumber\\
Y_{2,0}^1(B)&=0\, , \quad  Y_{2,1}^1(B)=B\, , \quad Y_{2,2}^1(B)=2B\, ,
\end{eqnarray}
which are in agreement with (\ref{eq:formula1}).

\noindent
(ii) One can also verify that for general $b$:
\begin{equation}
Y_{b,0}^1(B)=0\, , \quad Y_{b,1}^1(B)=B^{b-1}\, ,
\end{equation}
which follows from the first two equations in (\ref{eq:rec1}) and is again in agreement with (\ref{eq:formula1}).

\noindent
(iii) Suppose that (\ref{eq:formula1}) is valid for $b\leq m$ and every possible $i$, as well as for $b=m+1$ and $i\leq j<m$. Then, we will prove that (\ref{eq:formula1}) is valid for $b=m+1$ and $i=j+1$. To this end we use the transformation rules that follows directly from (\ref{eq:rec1}):
\begin{eqnarray}
Y_{m+1,0}^1(B)&=&BY_{m,0}^1(B),\nonumber \\
Y_{m+1,1}^1(B)&=&BY_{m,0}^1(B)+BY_{m,1}^1(B),\nonumber \\
Y_{m+1,i}^1(B)&=&\sum_{k=0}^{i-1}B^{i-k}Y_{m,k}^1(B)+BY_{m,i}^1(B), \qquad \mathrm{for} \quad 2\leq i\leq m, \nonumber \\
Y_{m+1,m+1}^1(B)&=&\sum_{k=0}^mB^{m+1-k}Y_{m,k}^1(B)+Y_{m,m}(B,B,B), \label{eq:recur2}
\end{eqnarray}
From (\ref{eq:recur2}) we obtain
\begin{equation}
Y_{m+1,j+1}^1=BY_{m+1,j}^1+BY_{m,j+1}^1+B(1-B)Y_{m,j}^1\, .
\end{equation}
Using (\ref{eq:formula1}) for $Y_{m+1,j}^1$, $Y_{m,j+1}^1$, and $Y_{m,j}^1$ it is a matter of
straightforward algebra to show that
\begin{equation}
Y_{m+1,j+1}^1(B)=\sum_{k=0}^{j-1}{{m-j+2}\over{k+1}}{{m+1}\choose{k}}
{{j-1}\choose k} B^{m+j-k-1}\, .
\end{equation}

\noindent
(iv) If (\ref{eq:formula1}) is valid for $b\leq m$ with every possible $i$, and for $b=m+1$ with
$i\leq m$, then from (\ref{eq:recur2}) one has
\begin{equation}
Y_{m+1,m+1}^1(B)=BY_{m+1,m}^1(B)+B(1-B)Y_{m,m}^1(B)+X_{m,m}(B)\, .
\end{equation}
Again, inserting (\ref{eq:formula1}) for $Y_{m+1,m}^1(B)$ and $Y_{m,m}^1(B)$, together with
$X_{m,m}(B)$ (see equation (\ref{eq:Z2})) one finds
\begin{equation}
Y_{m+1,m+1}^1(B)=\sum_{k=0}^{m-1}{2\over{k+1}}{{m+1}\choose k}{{m-1}\choose k}B^{2m-k-1}\, ,
\end{equation}
by which equation (\ref{eq:formula1}) and also (\ref{eq:a11}) has been proven for general $b$ (since $A_{11}(b)=Y_{b,b}^1(B_b^*)$).

The derivation of equation (\ref{eq:a12}) can be performed in a similar way, which we are not going to elaborate here.

\section{\label{app:drugi}}

The objective of this appendix is to derive equation (\ref{eq:a21}). Putting (\ref{eq:Z2}) into equation (\ref{eq:Z1}) one immediately obtains
\begin{equation}
\fl A_{21}(b)=(b-1)(B_b^*)^{2(b-1)}+(B_b^*)^{b-1}\sum_{i=1}^{b-2}\sum_{k=i}^{b-2}{{i(i+1)}\over{k+1}}
{{b-1}\choose k}
{{b-i-2}\choose{k-i}}(B_b^*)^k \, .\label{eq:Z3}
\end{equation}
Introducing label $f(i,k)$ with
\begin{equation}
f(i,k)={{i(i+1)}\over{k+1}}{{b-1}\choose k}
{{b-i-2}\choose{k-i}}(B_b^*)^k,
\end{equation}
the double sum in (\ref{eq:Z3}) can be rewritten as
\begin{eqnarray}
 \sum_{i=1}^{b-2}\sum_{k=i}^{b-2}f(i,k)=f(1,1)+f(1,2)+&\cdots&+f(1,b-2)\nonumber\\
  {\hbox{\hskip4cm}} + f(2,2)+&\cdots&+f(2,b-2)\nonumber\\
&\vdots&\nonumber\\
&&+f(b-2,b-2)\nonumber
\end{eqnarray}
\begin{equation}
=\sum_{i=1}^{b-2}\sum_{k=1}^if(k,i)=\sum_{i=1}^{b-2}{1\over{i+1}}{{b-1}\choose i}
(B_b^*)^i
\sum_{k=1}^i k(k+1){{b-k-2}\choose{i-k}},
\end{equation}
and, since
\begin{equation}
\sum_{k=1}^ik(k+1){{b-k-2}\choose{i-k}}=2{{b}\choose {i-1}},
\end{equation}
which can easily be proved by induction, we finally get (\ref{eq:a21}).

\section{\label{app:treci}}
For the adsorption of directed walks on homogeneous square lattice \cite{Privman} it was found
that implicit equation
\begin{equation}
w(1-x^2)(1+x-xw)=1 \label{eq:Privman}
\end{equation}
defines the value $x_\infty(w)$ for which $\langle N\rangle \to\infty$, when $w={\rm exp}(-\varepsilon_w/k_BT)>w_c$, where
$\langle N\rangle$ is the average lenght of the walk ($x$ is statistical weight of a step, 
whereas  $\varepsilon_w<0$ is an energy assigned to each step at the surface). Values $x=x_c=\sqrt 2-1$ and $w=w_c=1+1/\sqrt 2$ correspond to the adsorption transition. From
\eref{eq:Privman} straightforwardly follows that 
\begin{equation}
w-w_c\sim (x_\infty(w)-x_c)^{1/2}\, , \label{eq:sing}
\end{equation}
for $w\to w_c^+$ and $x\to x_\infty^-(w)$. It was also shown in \cite{Privman} that 
\[
\langle N\rangle\sim {1\over{x_\infty(w)-x}}\quad {\mathrm and} \quad
\langle M\rangle\sim {1\over{ x_\infty(w)-x}}{{\d x_\infty (w)}\over{\d w}}\, ,
\]
where $\langle M\rangle$ is the average number of steps at the surface. Then, by the use of
\eref{eq:sing}, one easily obtains
\[
\langle M\rangle \sim \langle N\rangle^{1/2}\, ,
\]
which means that crossover exponent $\phi$ is equal to 1/2 for directed walk adsorption on 2d Euclidean lattices.

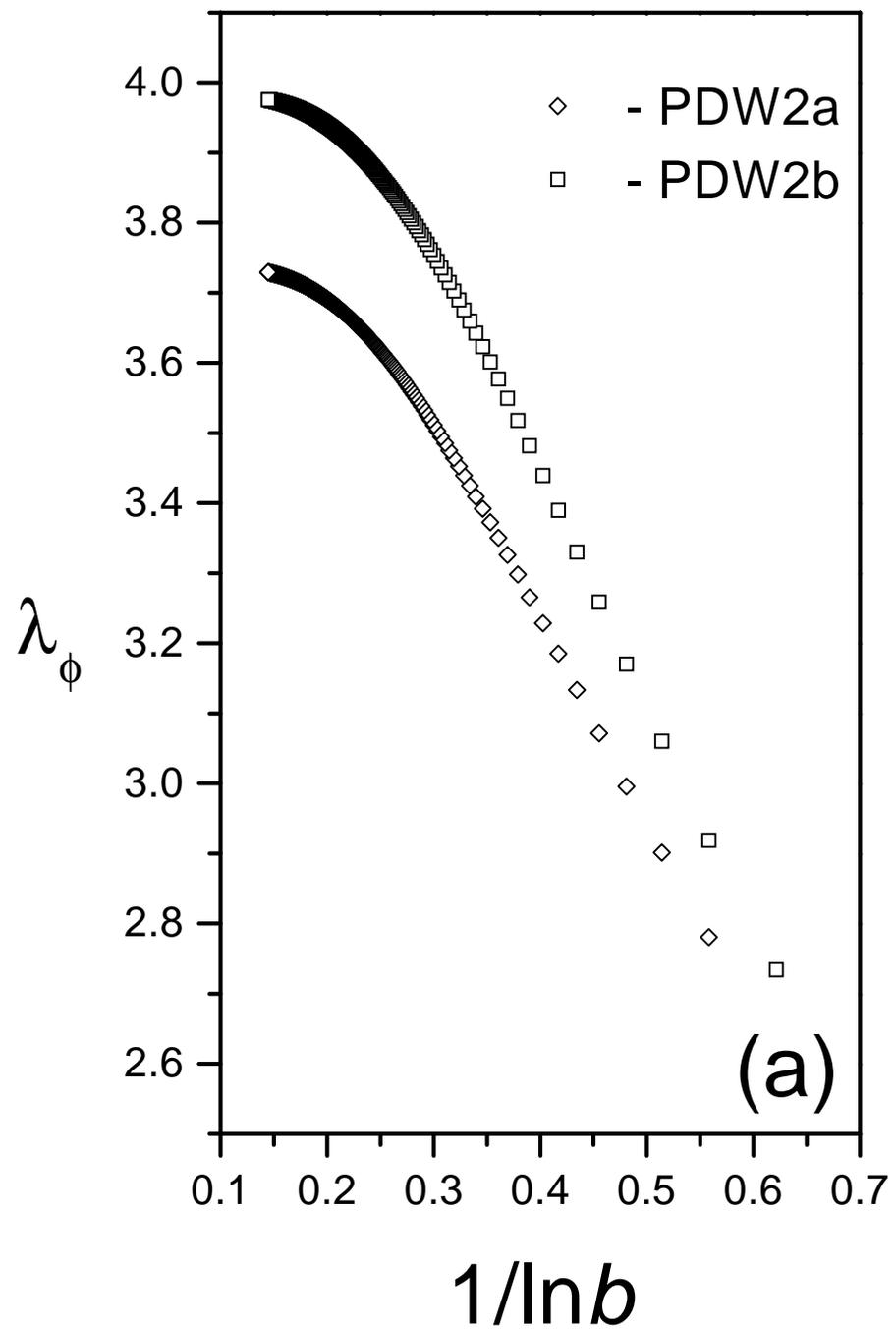
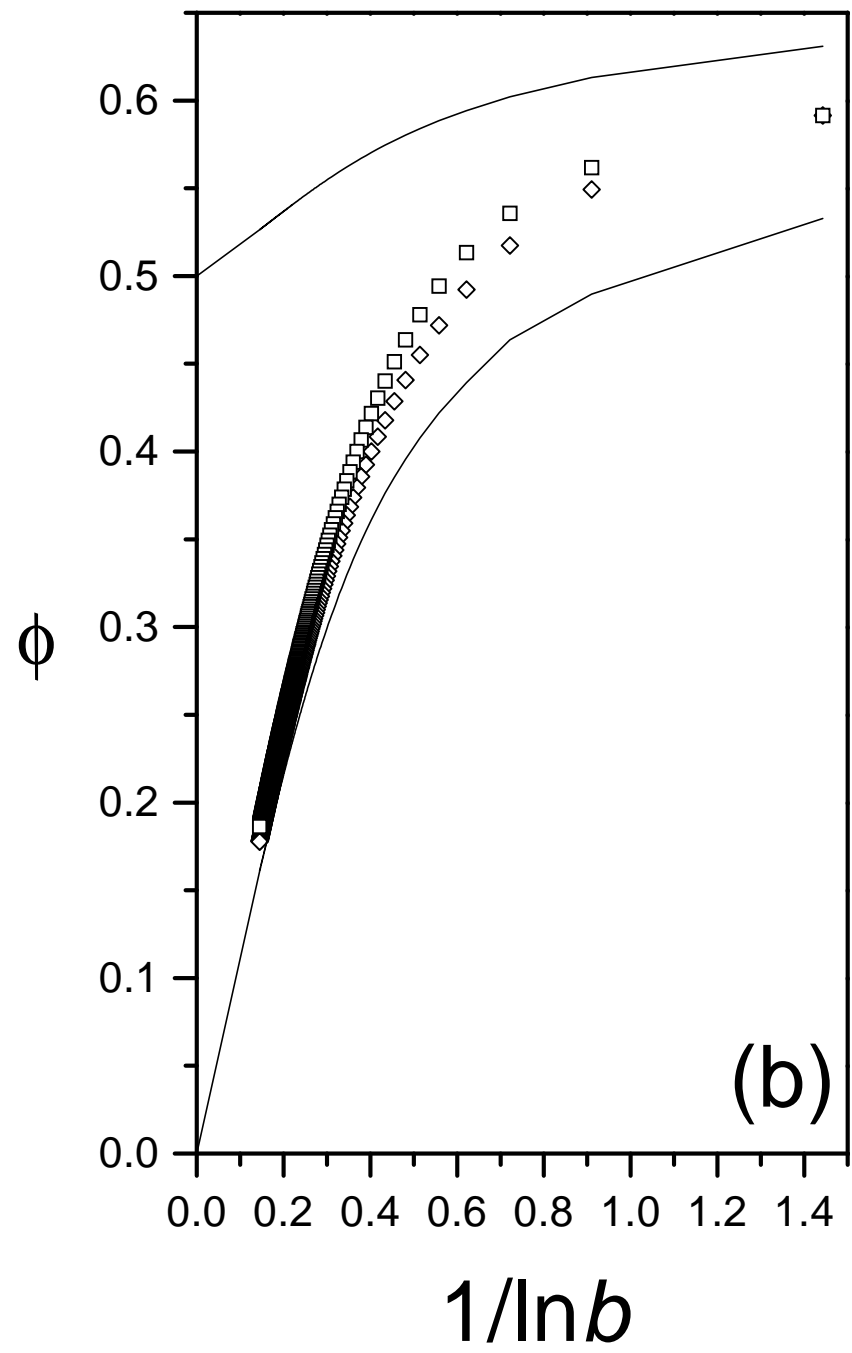
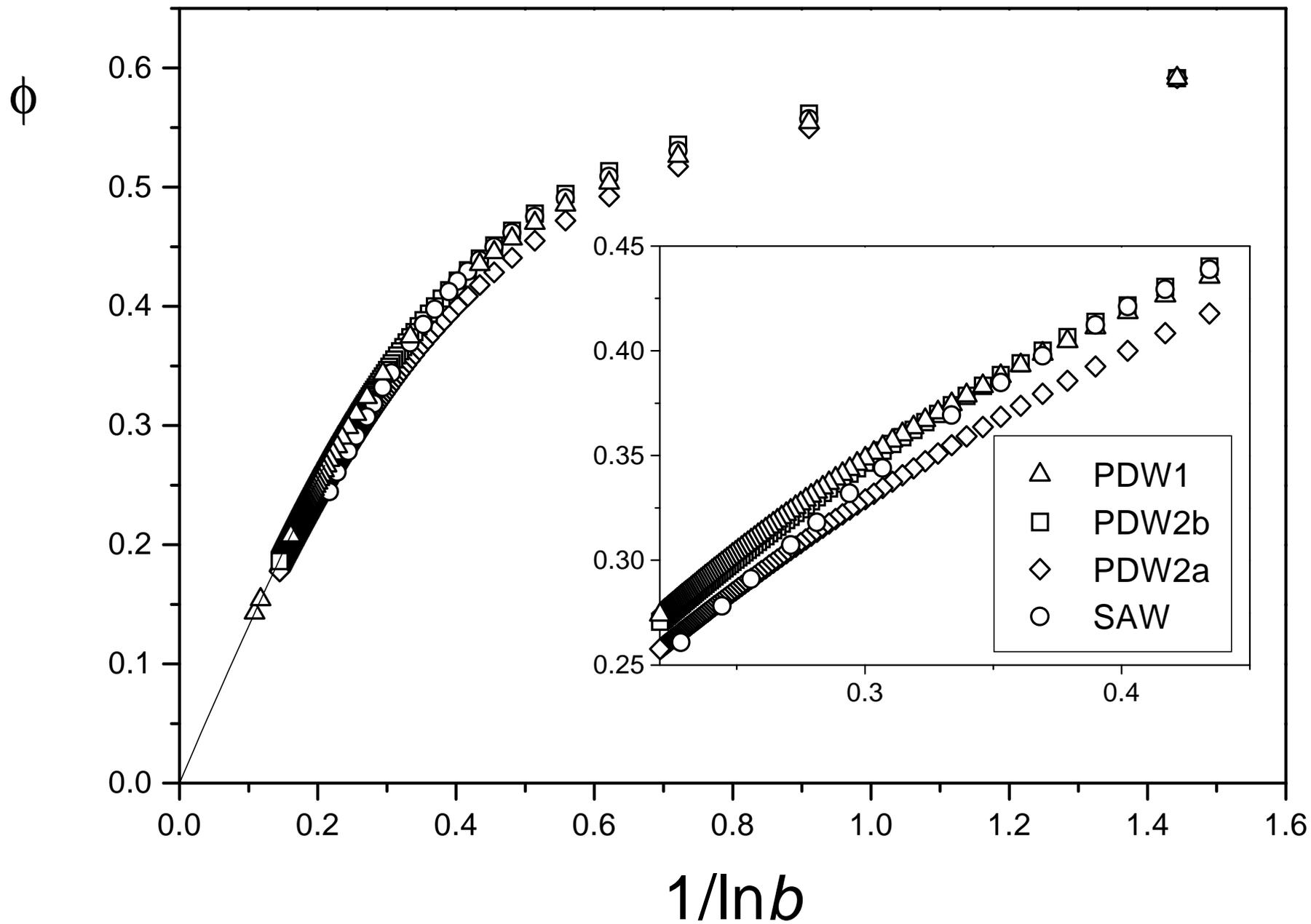
\end{document}